# Analysis of critical points of the In-Vessel Retention safety evaluation

**L. Carénini, F. Fichot**

IRSN, BP3 – 13115 St Paul lez Durance, France

laure.carenini@irsn.fr, florian.fichot@irsn.fr

I. Introduction

In-Vessel Retention (IVR) strategy for nuclear reactors in case of a Severe Accident (SA) intends to stabilize and retain the corium in the vessel by using the vessel wall as a heat exchanger with an external water loop. This strategy relies on simple actions to be passively taken as soon as SA signal is raised: vessel depressurization and reactor pit flooding. Then, the strategy is successful if the vessel keeps it integrity, which means that the heat flux coming from the corium pool does not exceed the cooling capacity of the External Reactor Vessel Cooling (ERVC) at each location along the vessel wall (no vessel melt-through) and the ablated vessel wall is mechanically resistant. The main uncertainties in this IVR safety evaluation are associated to the thermal load applied from the corium pool to the vessel wall and the resulting minimum vessel thickness after ablation. Indeed, the heat fluxes distribution along the vessel wall is directly dependent on the corium stratification which occurs as a result of thermochemical interactions in the pool: when liquid steel is mixed with $UO_2$ and partially oxidized Zr coming from the degradation of the fuel and claddings, there is a phase separation between oxide and metal phases due to a gap of miscibility. The primordial impact of the corium behaviour in the lower plenum of the reactor vessel on the IVR safety evaluation was clearly highlighted in the Phenomena Identification Ranking Table (PIRT) on IVR performed in the frame of the European IVMR (In-Vessel Melt Retention) project (Fichot et al., 2019). As a result, the focus is made in this paper on the critical points which impact the value of the minimum vessel thickness or equivalently the maximum heat flux reached at the outer surface of the vessel wall. Efficiency of the ERVC and mechanical resistance of the vessel wall are consequently not discussed here.

The main objective is to identify the generic critical situations leading to an excessive heat flux to the vessel wall and the investigation of possible means to avoid them. In this perspective, the calculations of IVR strategy done by the project partners for different reactor designs and accident scenarios were used as a database to identify and understand the occurrence of critical configurations with excessive heat flux to the vessel wall. The results of 25 sequences are used, which correspond to 9 different reactor designs: a generic PWR 900MWe, a PWR 1100MWe with heavy reflector, a generic PWR 1300MWe, a generic Konvoi 1300MWe, a generic German BWR69, a Nordic BWR, a BWR-5 Mark II, a VVER1000 and a VVER440/v213. In addition, different SA integral codes (ASTEC, ATHLET-CD, MAAP -combined with MAAP_EDF and PROCOR codes for simulation of lower plenum behavior-, MELCOR and RELAP/SCDAPSIM codes) are used.

II. Results of reactor calculations: role of the reactor power and of the SA scenario

The minimum vessel thicknesses evaluated in each reactor calculation are synthetized in Figure 1 and vary from 11cm to less than 1cm.

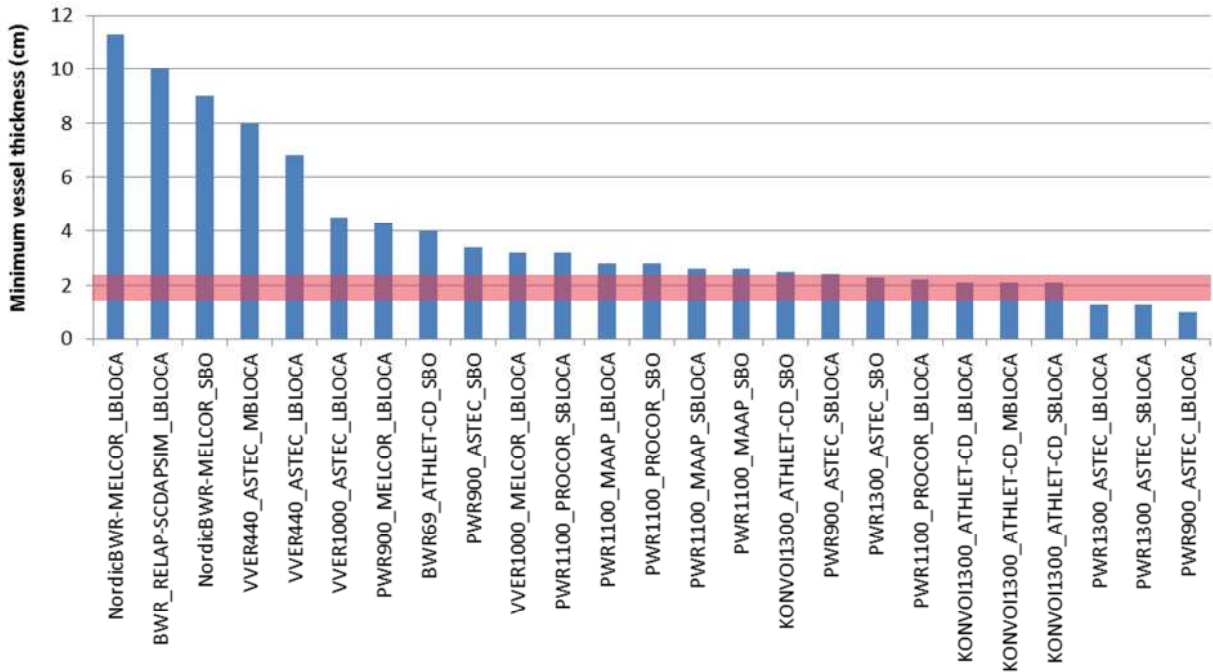

*Figure 1: Results of the minimum vessel thickness evaluated for each reactor calculation[1]*

Corresponding results for the maximum heat flux reached in transient situation at the outer surface of the vessel wall, vary between 0.35 to 3.5MW/m². It is worth noting that only 3 calculations lead to vessel thickness below 2cm and maximum outer heat flux above 2MW/m²: LBLOCA scenario for the generic PWR900 and PWR1300 reactors and SBLOCA scenario for the generic PWR1300 reactor. A general trend of increase of the maximum heat flux with the electric power of the reactor is visible in Figure 2. It confirms the intuitive relation between the power of the reactor and the intensity of the thermal load on the vessel wall. However, for a given nominal power, a large dispersion of the maximum heat flux depending on the reactor design is obtained, illustrating that **applicability of IVR strategy cannot be determined only from the reactor power**.

---

[1] Since simulations deal with SA here, for each LOCA scenario, unavailability of active injection systems is also assumed in addition to the primary break.

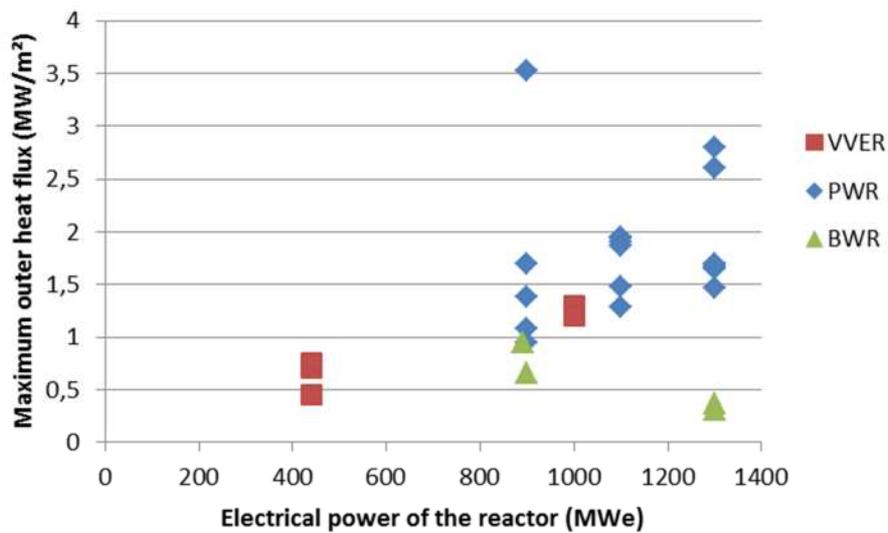

*Figure 2: Maximum outer heat flux depending on reactor design and power*

With respect to the scenario, the results of SA simulations show that the LBLOCA scenario with loss of off-site and in-site electrical supply (Station black out – SBO) is the most critical scenario for all nuclear power plant designs, as expected because of the high level of residual power. However, for other scenarios like SBO or SBLOCA, critical situations may also be reached if no mitigation measure is considered to significantly delay the beginning of the core degradation. For the LBLOCA+SBO scenario, such mitigation measures are not possible since core degradation occurs in few minutes. Consequently, due to its fast kinetics, this scenario appears to be the most difficult to deal with for IVR. Some design provisions can make the LBLOCA less critical, in particular if water is available to delay the time of complete dry-out of the vessel. However, such provisions, necessary shortly after the initiating event, have to be fully passive. Additionally, for efficiency, water shall not be lost at the RCS breach.

The results presented in Figure 3 illustrate that the time before reaching the maximum heat flux at the outer surface of the vessel is one of the main parameters: the longer it is, the lower the residual power to be extracted. Indeed, the decay heat 10h after scram is divided by 2 compared to the value 1h after scram. Therefore, there is a critical time window during which, if the vessel lower plenum remains filled with water, the IVR strategy can be successful even for high power reactors. As a consequence, **a high power reactor is not necessarily less suitable for IVR compared to a reactor with a smaller power if the amount of available water reaching the vessel is also higher and leads to sufficiently delay the complete dry-out of the vessel, allowing a decrease of the decay heat during the first hours after the reactor scram.** This point is discussed in more details in the next part.

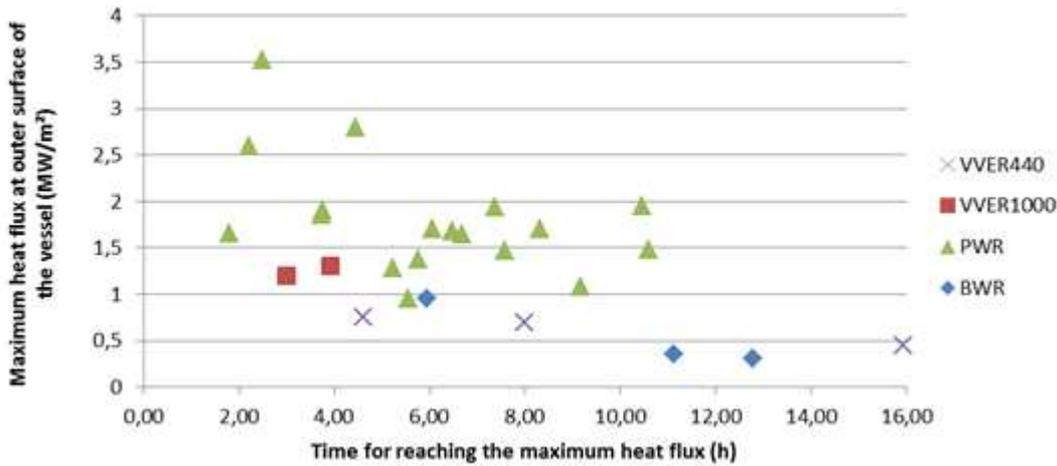

*Figure 3: Maximum heat flux in the lower head depending on time necessary to reach it*

III. The role of primary water volume and simultaneous injection

In this part a simple analysis is proposed to better understand the importance of the reactor power and of the design.

First, the residual power per unit mass of UO$_2$ is approximated as:

$$P(t) = P(\tau_0)\exp\left(-\frac{t - \tau_0}{\tau_p}\right) \quad \text{Eq. 1}$$

where $P(\tau_0) \sim 250$ W/kg, $\tau_0 = 1$h and $\tau_p = 13$h

We assume that there is a sufficient residual mass of water $M_w$ in the lower plenum. The time $\tau_{dry-out}$ which is necessary to evaporate that water is given by:

$$\int_{\tau_0}^{\tau_{dry-out}} kM_{UO2}P(t)dt = M_w \Delta h_{vap} \quad \text{Eq. 2}$$

where k is the fraction of power actually used for evaporation (typically of the order of 0.5, if we assume that water is above the molten pool) and $\Delta h_{vap}$ is the massic enthalpy of vaporization of water.

The integration gives:

$$\tau_{dry-out} = \tau_0 - \tau_p \ln\left(1 - \frac{M_w \Delta h_{vap}}{kM_{UO2}P(\tau_0)\tau_p}\right) \quad \text{Eq. 3}$$

and the residual power at the instant of complete dry-out of the vessel is:

$$M_{UO2}P(\tau_{dry-out}) = M_{UO2}P(\tau_0)\left(1 - \frac{M_w \Delta h_{vap}}{kM_{UO2}P(\tau_0)\tau_p}\right) \quad \text{Eq. 4}$$

This relation shows that, if we compare, for two different reactor designs, the residual power at the time of maximum heat flux, it is not just directly proportional to the mass of fuel. It is also multiplied by a factor that increases with the mass of fuel, because of the reduced time to reach complete dry-out of the vessel. For example, comparing a 1000MWe to a 500MWe reactor, and considering that the mass of water is the same ($M_w = 50t$) and the time of corium relocation in the lower plenum is also the same ($t \sim 1h$), the residual power at the time of maximum heat flux is 3 times higher (instead of 2 times as intuitively estimated) for the 1000MWe case.

It is also interesting to look at the mass of water that should be added to get the same level of residual power at the time of the vessel dry-out, compared to a reference ($P_{ref}$). The mass of water should be increased according to the following formula:

$$M_W(\alpha . P_{ref}) - M_W(P_{ref}) = (\alpha - 1)\frac{k\, M_{UO2}(P_{ref})P(\tau_0)\tau_p}{\Delta h_{vap}} \qquad \text{Eq. 5}$$

Considering the VVER440 reactor as a reference, we estimate the increase of the mass of water available in the vessel, which will be necessary in the case of a higher power reactor. In order to get the same residual power level as in a 440MWe at dry-out, it would be necessary to keep the lower plenum under water for 18h for a 1000MWe reactor, which means adding approximately 165t of water, thanks to passive or active systems. It is interesting to notice that the corresponding average flow rate 3.1kg/s remains limited.

| Reactor power | Reference case 440MWe | 1000MWe | 1300MWe |
|---|---|---|---|
| **Additional water mass** | - | 165 t | 253 t |
| **Time before dry-out** | 7h15 | 18h (3h20 without injection) | 21h (2h50 without injection) |
| **Water flow rate (average) needed for in vessel injection** | - | 3.1 kg/s | 3.8 kg/s |

Even if that total amount of water cannot be injected, for high power reactors, simultaneous in-vessel injection appears as a relevant action to reduce the intensity of focusing effect and the maximum heat flux.

IV. The role of the metallic structures

The third main contributor is the mass of molten steel incorporated in the molten pool at the time when the maximum heat flux is reached. Obviously, increasing the mass of molten steel incorporated in the pool reduces the focusing effect. However, an accurate evaluation of this mass requires being able to simulate transient configurations with progressive incorporation in the pool. It is proposed to use the variable $X_{focus}$ ($= UO_2\, mass/molten\, steel\, mass$), as a simple and meaningful parameter to compare reactor designs. But it has to be compared at different times in order to take into account the transient effects. Indeed, the mass of molten steel increases with time (at least due to the vessel wall melting) and also the mass of UO$_2$ may increase if only part of the core is relocated first. These temporal evolutions may have an impact on the evaluation of this $X_{focus}$ ratio leading to different conclusions regarding the possibility to reach critical configurations for IVR in a specific design. In Figure 4, this mass ratio is evaluated at the time when the maximum heat flux is reached and also at the time of "steady-state" (i.e. no more material relocation in the lower head) for the different reactor calculations.

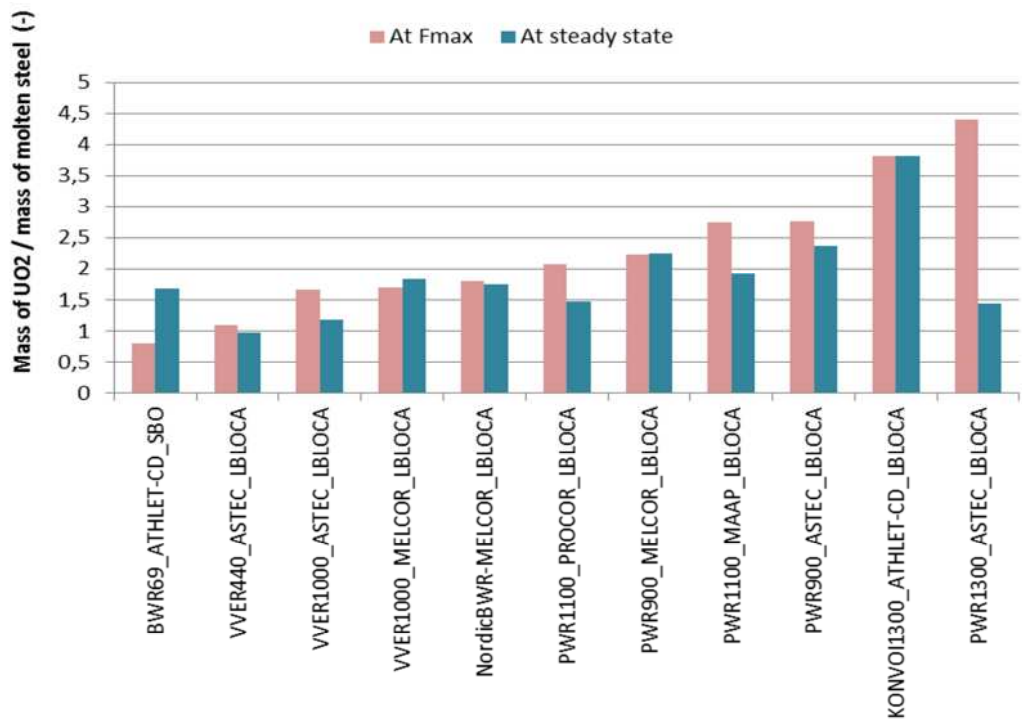

*Figure 4: Ratio of UO$_2$ mass to molten steel in the lower head at the time when the maximum heat flux is reached and at steady-state*

In most of the cases, the ratio increases when it is calculated at the time of the maximum heat flux in transient configurations. This increase is mainly associated to the lower mass of molten steel compared to the one evaluated at steady state. In particular for the PWR1300 design, this effect is strong and should be associated to the fact that the mass of the core support plate is not incorporated yet in the pool when the maximum heat flux is reached. Conversely, for the BWR69 reactor with ATHLET-CD, the ratio significantly increases at steady state due to the fact that at maximum heat flux, only part of the core is relocated in the lower head.

As a consequence, analysis of calculations highlighted the importance of the consideration of temporal aspects even when looking at simple parameters. To include the possible consequences of this transient evolution and have a first conservative evaluation, it is then recommended to consider the total mass of UO$_2$ in the core but only the mass of molten steel which is likely to be incorporated in the lower head due to its location (structures in the lower head for example). **Assumptions on the mass of steel coming from the core part during degradation and from the core support plate (when not submerged in the corium pool) should be considered with caution and it is recommended to perform more detailed analysis to evaluate the range of variation of this available molten steel mass.**

V. Mechanisms leading to transient peak heat fluxes

The critical scenarios leading to transient "peak" heat flux are identified as being, in most of the cases, configurations with reduced top metal layer and resulting strong focusing effect with high heat flux (around or above 2MW/m²). The maximum heat flux is mainly identified in transient situation, with an increase factor up to 3 compared to the one obtained in stable configuration and above 1.5 in half of the calculations. It corresponds mainly to 3-layers configurations (i.e. with heavy metal layer formation) with a top a metal layer thickness between 10-20cm (cf. Figure 5).

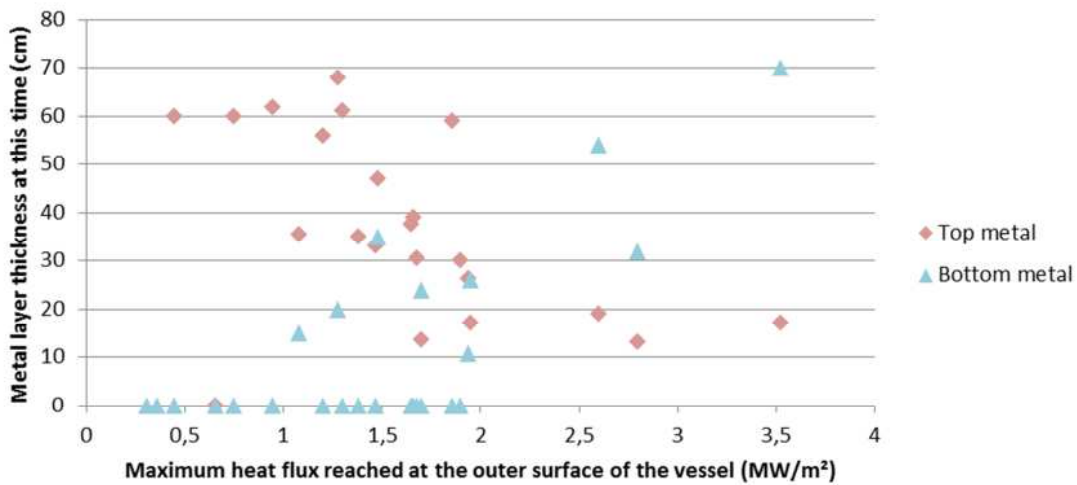

*Figure 5: Top and bottom metal layers thicknesses at the time when maximum heat flux is reached*

In addition, the consequences of the metal superheat after the stratification inversion were identified as another critical scenario. This last scenario appears to be less frequent than the first one, due to possible presence of metal at a lower temperature already at the top of the oxide pool when stratification inversion occurs, preventing an abrupt increase of the average metal temperature. However, in both cases, **the formation of a metal layer initially heavier than the oxide leads to potentially critical configurations.** Importance of a high oxidation degree of the corium pool for preventing such formation of large heavy metal layer is illustrated in Figure 6, using Salay-Fichot model (Salay & Fichot, 2005) for the evaluation of metal and oxide phases composition at equilibrium.

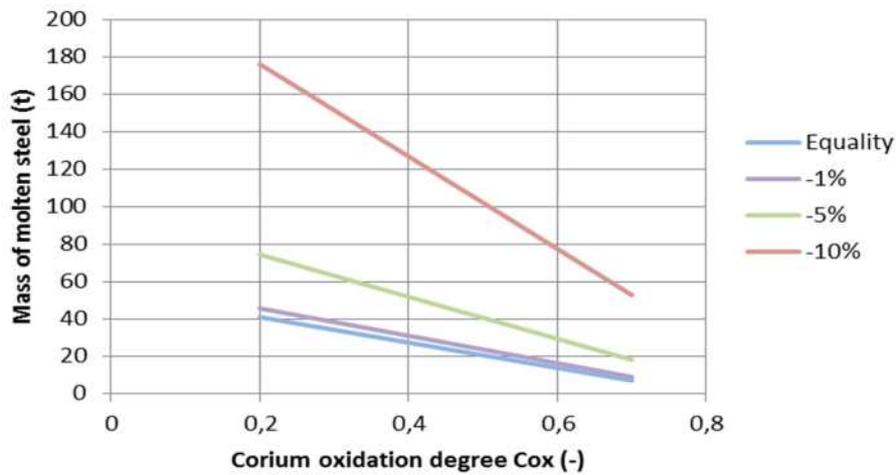

*Figure 6: Mass of molten steel needed to form a metal phase lighter than the oxide depending on the corium oxidation degree and on the necessary density difference with the oxide for triggering the stratification inversion - Typical 1000MWe reactor ($UO_2$ mass of 86t, U/Zr=1.4, oxide density taken at 8000kg/$m^3$)*

In addition, results of reactor calculations show that increasing the time before molten pool formation in the lower plenum increases the oxidation degree (cf. Figure 7).

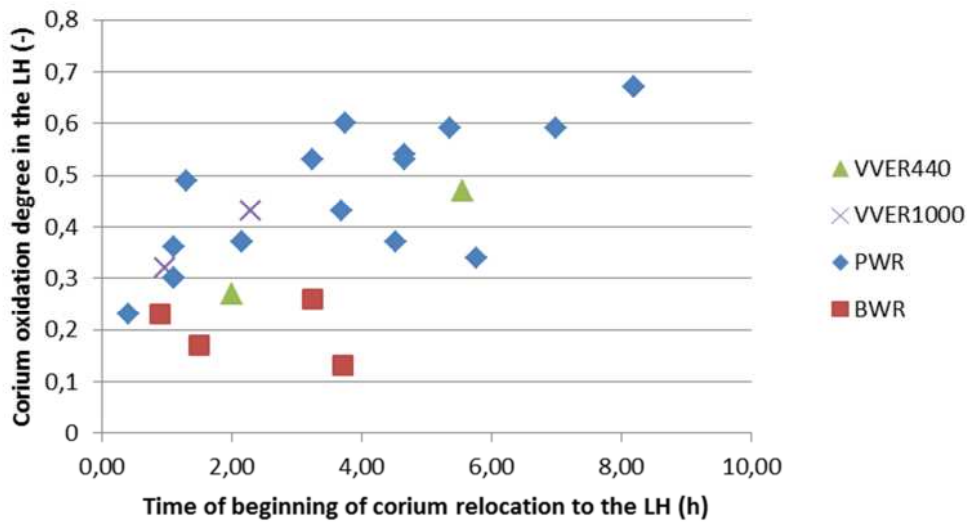

*Figure 7: Corium oxidation degree (moles of Oxidized Zirconium/Total moles of Zirconium) function of time of corium relocation to the lower head*

Consequently, increasing the time before molten pool formation in the lower plenum leads to two beneficial effects for the IVR strategy: residual power reduction and prevention of heavy metal layer formation. **Therefore, means to avoid fast kinetics scenarios play a primary role in the success of IVR strategy.**

VI. Conclusion

Thanks to the results of the reactor calculations, it was possible to illustrate the combined impact on the IVR safety evaluation of the design and of the scenario. The three main parameters impacting the maximum heat flux at the outer surface of the vessel or equivalently the minimum vessel thickness were identified as being: (i) the nominal power of the reactor, (ii) the means (i.e. water injection for example) available to delay the core dry-out and (iii) the mass of molten steel incorporated in the pool at the time of the maximum heat flux at the outer surface of the vessel. Thus, the time before reaching the maximum heat flux and the corresponding mass of molten steel incorporated in the pool at this time appear to be critical points of the IVR safety evaluation. For future reactor designs, preventing fast kinetics scenarios and assuring that a sufficient amount of metallic structures will be molten before reaching maximum heat flux in all circumstances (e.g. metallic structures immerged in the corium when it relocates in the lower head) looks as two of the most important points to examine in order to optimize the IVR strategy. For existing reactors, the correct evaluation of the kinetics of the mass of molten steel incorporated in the pool is an important point to prove the success of IVR.

Analysis of critical scenarios with excessive heat flux reveals that associated mechanisms are the thinning of the top metal layer due to transient 3-layers configuration which leads to a strong focusing effect and the dissipation of the metal superheat due to the stratification inversion between oxide and metal layers. If the second case appears to be less frequent than the first one, both are related to the formation of a metal layer initially heavier than the oxide. Importance of oxidation degree of the corium pool for preventing such critical situation was illustrated, showing that increasing the time before molten pool formation in the lower plenum also increases the oxidation degree and thus prevents the heavy metal layer formation. This strengthens the criticality of fast kinetics scenarios for IVR strategy evaluation.

More generally, the analysis performed based on the results of the reactor calculations and presented in this paper outlines the capabilities of integral (or dedicated) codes for transient evaluation of IVR. Not only they are able to provide more accurate evaluations of the maximum heat flux through the vessel and minimum residual vessel thickness but also they can be used to improve or optimize a design for IVR. It should however be kept in mind that, in the analysis, limitations depending on the level of maturity of the code used and also depending on remaining phenomenological uncertainties were considered, in relation with the work done in the IVMR project on code modelling (Carénini et al., 2019) and reactor calculations (Sangiorgi et al., 2019). Regarding the remaining uncertainties, it should be noted that they do not impact analysis of main mechanisms influencing the maximum heat flux as done in this report but mainly the evaluation of the maximum heat flux itself by impacting corresponding corium pool configuration. Actually, best estimate integral code simulations can only give a first estimate (which may be quite accurate already) but the analysis should be completed by sensitivity studies and if necessary more detailed analysis. In particular, it may be necessary to take into account the specificity of the design for the kinetics of corium arrival to the lower plenum and of molten steel addition. Both kinetics may depend on geometrical details which cannot be properly taken into account by generic codes.

Acknowledgements

Participants of IVMR project who provided results of reactor calculations used in this report are greatly acknowledged for their inputs:
- BAKOUTA Nikolai from EDF,
- EDERLI Stefano and MASCARI Fulvio from ENEA,
- LECOMTE Marylène and SAGAN Michael from Framatome,
- PANDAZIS Peter from GRS,
- JOBST Matthias from HZDR,
- CARENINI Laure from IRSN,
- BARNAK Miroslav and MATEJOVIC Peter from IVS,
- VILLANUEVA Walter and ZHANG Huimin from KTH,
- VALINČIUS Mindaugas and KALIATKA Algirdas from LEI,
- TECHY Zsolt and KOSTKA Pal from NUBIKI,
- THOMAS Robin from TRACTEBEL,
- KOTOUC Miroslav and VOKAC Petr from UJV.

This work was performed within the HORIZON 2020 IVMR Project No. 662157.